\documentclass{IEEEtran}  

\IEEEoverridecommandlockouts                              

\makeatletter
\renewcommand{\@IEEEsectpunct}{.\ \,}
\makeatother

\usepackage{graphics} 
\usepackage{epsfig} 
\usepackage{mathptmx} 
\usepackage{times} 
\usepackage{amsmath} 
\usepackage{amssymb}  
\usepackage[]{units}
\usepackage{gensymb}
\usepackage{placeins}
\usepackage{dblfloatfix}  

\title{\LARGE \bf
In-Ear Measurement of Blood Oxygen Saturation: An Ambulatory Tool Needed To Detect The Delayed Life-Threatening Hypoxaemia in COVID-19}

\author{Harry J. Davies, Ian Williams , Nicholas S. Peters, and Danilo P. Mandic
}

\begin{document}

\maketitle
\thispagestyle{empty}
\pagestyle{empty}

\begin{abstract}
Non-invasive ambulatory estimation of blood oxygen saturation has emerged as an important clinical requirement to detect hypoxemia in the delayed post-infective phase of COVID-19, where dangerous hypoxia may occur in the absence of subjective breathlessness. This immediate clinical driver, combined with the general quest for more personalised health data, means that pulse oximetry measurement of capillary oxygen saturation (SpO\textsubscript{2}) will likely expand into both the clinical and consumer market of wearable health technology in the near future. In this study, we set out to establish the feasibility of SpO\textsubscript{2} measurement from the ear canal as a convenient site for long term monitoring, and perform a comprehensive comparison with the right index finger -  the conventional clinical measurement site. During resting blood oxygen saturation estimation, we found a root mean square difference of 1.47\% between the two measurement sites, with a mean difference of 0.23\% higher SpO\textsubscript{2} in the right ear canal. Through the simultaneous recording of pulse oximetry from both the right ear canal and index finger during breath holds, we observe a substantial improvement in response time between the ear and finger that has a mean of 12.4 seconds and a range of 4.2 - 24.2 seconds across all subjects. Factors which influence this response time, termed SpO\textsubscript{2} delay, such as the sex of a subject are also explored. Furthermore, we examine the potential downsides of ear canal blood oxygen saturation measurement, namely the lower photoplethysmogram amplitude, and suggest ways to mitigate this disadvantage. These results are presented in conjunction with previously discovered benefits such as robustness to temperature, making the case for measurement of SpO\textsubscript{2} from the ear canal being both convenient and superior to conventional finger measurement sites for continuous non-intrusive long term monitoring in both clinical and everyday-life settings.

\end{abstract}


\section{Introduction}

\IEEEPARstart{O}{ne} of the major roles of blood is to supply oxygen to tissues throughout the body. This is achieved through the protein haemoglobin within red blood cells, which has a high affinity to oxygen. Thus, as blood passes through capillaries in the lungs, the haemoglobin in red blood cells binds to oxygen which is subsequently pumped through arteries via the heart and transported to various tissues. The maintenance of a high arterial blood oxygen saturation is therefore extremely important, as otherwise tissues cease to be adequately supplied. The term blood oxygen saturation specifically refers to the proportion of haemoglobin in the blood that is carrying oxygen \cite{Jubran1999}, and is given by

\begin{equation}
    \text{Oxygen Saturation} = \frac{HbO_{2}}{HbO_{2} + Hb}
\end{equation}
where $Hb$ refers to haemoglobin not bound with oxygen and $HbO_{2}$ refers to haemoglobin bound to oxygen.
\\

Arterial blood oxygen saturation is typically estimated using pulse oximetry, with peripheral capillary oxygen saturation (SpO\textsubscript{2}) serving as a proxy, given that it is recorded on the body surface. It has been established that those with a healthy respiratory system typically exhibit SpO\textsubscript{2} values of 96-98\% at sea level \cite{Smith2012}. According to the World Health Organisation, hypoxia is defined as a blood oxygen saturation level of less than 94\%, while a blood oxygen level of less than 90\% may indicate the need for clinical action \cite{WorldHealthOrganization2011}. Hypoxic SpO\textsubscript{2} readings are a sign of hypoxia without breathlessness in COVID-19 patients \cite{Gattinoni2020a} \cite{Cascella2020} where the major respiratory failure peaks 10 days after initial infection \cite{Pan2020}. Indeed, in the case of COVID-19 it is strongly recommended that patients receive supplemental oxygen if their SpO\textsubscript{2} reading falls below 90\% \cite{Alhazzani2020}. Hypoxic SpO\textsubscript{2} levels may also occur in other examples of respiratory failure \cite{Khemani2012} and during breathing obstruction which is common in sleep apnea \cite{Oksenberg2010}. 

In practice, the SpO\textsubscript{2} levels are calculated using photoplethysmography (PPG), the non-invasive measurement of light absorption (usually green, red and infrared) through the blood. In short, when more blood is present, more light is reflected, so given a pulsatile increase in blood volume with each heart beat, PPG effectively measures the pulse. Depending on the level of blood oxygen saturation, the PPG measurements also experience a change in the ratio of light absorbance between the red and infrared light. Namely, the extinction coefficient of oxygenated haemoglobin with red light ($\approx 660nm$) is lower than it is for deoxygenated haemoglobin, and the reverse is true for infrared light ($ \approx 880-940nm$) \cite{Nitzan2014}. Thus, a simultaneous measurement of the absorbance of both infrared and red light allows for an estimation of blood oxygen saturation. Calculation of blood oxygen saturation, referred to as the \textit{ratio of ratios} method, will be fully explained in the Methods section. 

Reflective PPG can be measured from any site with skin that has vasculature \cite{Nilsson2007}, but is, for convenience, commonly measured from the fingers (usually the index finger), wrist and earlobe. This convenience comes with some disadvantages pertaining to 24/7 continuous measurements, which include the intrusive nature of measurements given that living life with a finger or earlobe clip is problematic, and also poor sensor skin contact and motion artefacts in the case of the wrist. On the other hand, the recent interest in the development of Hearables \cite{Goverdovsky2017}, has promoted the ear canal as a preferred site for the measurement of vital signs in eHealth technology. Indeed, the ear canal represents a unique opportunity for physiological measurements, due to its proximity to the brain (ear-EEG \cite{Looney2012}), the property of the ear canal to act as a shield from external electrical noise (nature-built Faraday cage \cite{Goverdovsky2017}), and the general fixed position of the head wrt. vital signs and neural function, as the head does not move much in daily life unlike sites such as the wrist. Owing to these desirable properties, the ear-canal has been established as a feasible wearable site by Ear-EEG \cite{Looney2012} \cite{Nakamura2020} and Ear-ECG \cite{Goverdovsky2017}.

When it comes to ear-PPG, unlike the finger PPG signal and the earlobe signal, it has been shown that the ear canal offers a photoplethysmogram which is stable and resistant to changes in blood volume which occur during hypothermia \cite{Budidha2015}. This is because peripheral areas of the body experience restricted blood flow during the cold, whereas the ear canal, being a narrow cavity surrounded by skin, maintains internal blood flow levels. Additionally, this makes the ear canal a preferred site for accurate core body temperature measurement. The PPG from the ear canal has also been shown to be far more sensitive than earlobe and finger PPG to amplitude variations that arise from respiration, thus allowing for a better measurement of respiration rate \cite{Budidha2018}. Furthermore, a significant delay has been evidenced between earlobe pulse oximetry and pulse oximetry on the hand or the foot for detection of hypoxemia (low levels of blood oxygen) \cite{Hamber1999}. The robustness of the PPG signal from the ear canal, the faster SpO\textsubscript{2} response time of the earlobe to changes in blood oxygen levels and the potential for relatively non-intrusive 24/7 ambulatory monitoring of COVID-19 patients form the motivation for our research into ear canal pulse oximetry. The aim of this study is therefore to investigate the feasibility, establish technical characteristics, and identify advantages in the recording of SpO\textsubscript{2} from the ear canal, compared with the most commonly used site, the right index finger. 

\section{Methods}

\subsection{Hardware}

\begin{figure}[h]
\centerline{\includegraphics[width=0.5\textwidth]{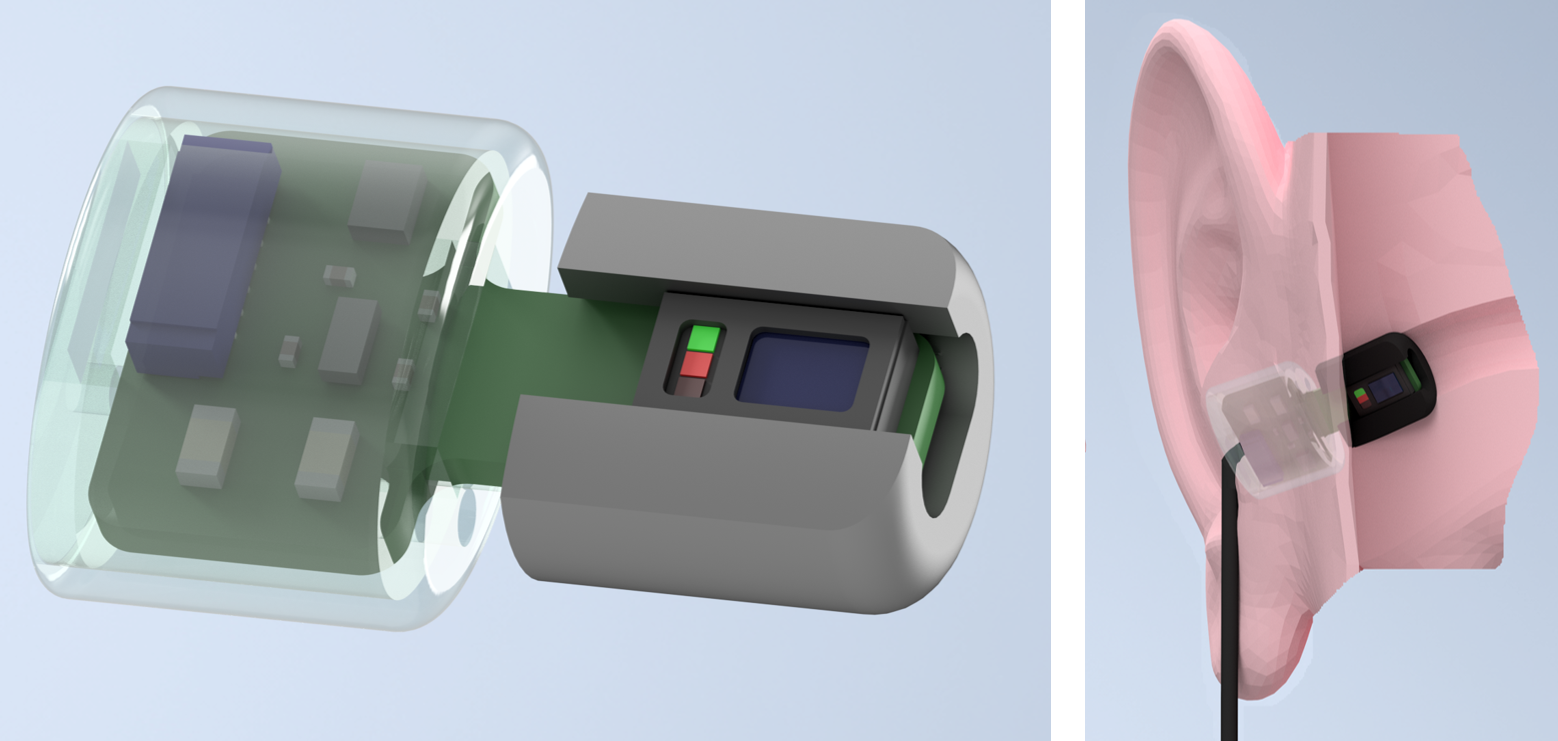}}
\caption{The in-ear photoplethysmography sensor used in our study. Left: Wearable in-ear photoplethysmography sensor for pulse oximetry, with a trimmed memory foam earbud and a 3D printed case to house the circuitry. Right: The sensor placed within the ear canal.}
\label{eareeg_3Dmodel}
\end{figure}

The photoplethysmography sensor used was the MAX30101 digital PPG chip by Maxim Integrated, consisting of green (537nm), red (660nm) and infrared (880nm) light emitting diodes as well as a photo-diode to measure the reflected light. In our study, the PPG sensor was embedded in a cut out rectangular section of a viscoelastic foam earbud \cite{Goverdovsky2016}, allowing for comfortable insertion. The three light wavelengths (at 537nm, 660nm, 880nm) gave us three measures of pulse, and also the signals required to estimate blood oxygen saturation and respiration rate. The PPG circuitry was packaged in a small 3D printed case which was positioned just outside the ear canal, as shown in Fig.~\ref{eareeg_3Dmodel}. This same sensor was secured to the right index finger with medical tape for the simultaneous finger PPG measurement.

\subsection{Experimental Protocol}

\begin{figure}[h]
\centerline{\includegraphics[width=0.38\textwidth]{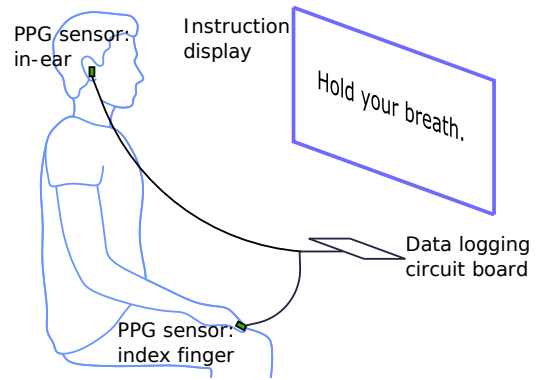}}
\caption{Simultaneous PPG recording from the ear canal and finger. One PPG sensor is placed in the right ear canal, and another is attached to the right index finger. The simultaneous recordings are linked via a circuit board which logs the respective data streams on SD cards. The subjects are instructed on when to breathe normally, exhale and hold their breath via a video played on a monitor (see Fig.~\ref{flow_digram_experiment} for the protocol), with a built in count down so that they also know how long they have left in each phase.}
\label{recording_diagram}
\end{figure}

\begin{figure}[h]
\centerline{\includegraphics[width=0.28\textwidth]{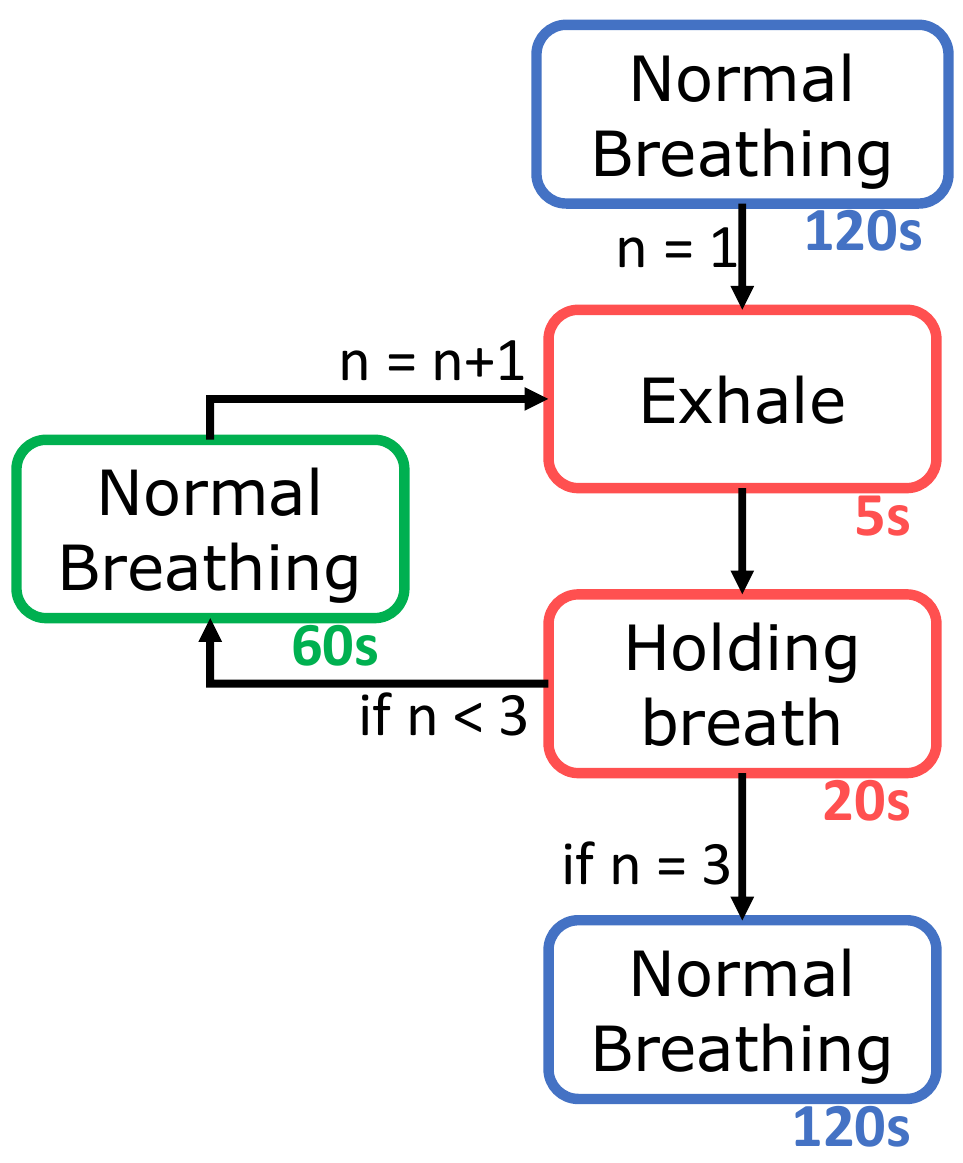}}
\caption{Flow diagram of the experiment, outlining the sequence of participants' breathing. The duration of each stage is provided beneath in seconds. The exhale and holding breath stages are repeated three times, the 60 seconds normal breathing is repeated twice, and 120 seconds of normal breathing occurs both at the start and at the end of the experiment.}
\label{flow_digram_experiment}
\end{figure}

\begin{figure*}[ht]
\centerline{\includegraphics[width=0.85\textwidth]{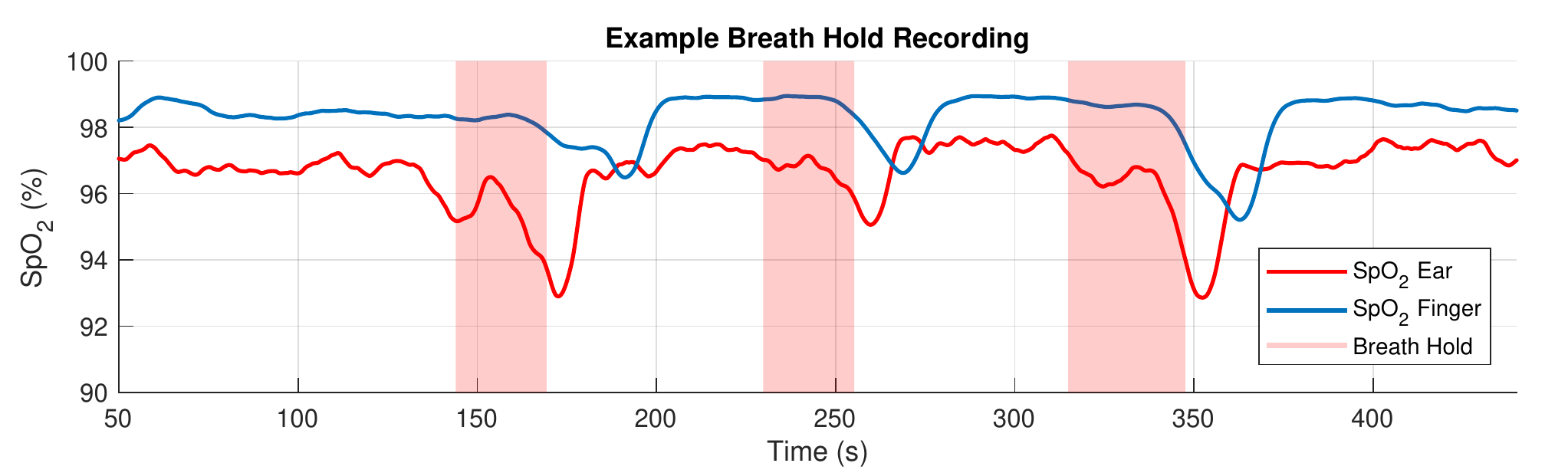}}
\caption{Exemplar SpO\textsubscript{2} recording following the breath hold protocol (see Fig.~\ref{flow_digram_experiment}). The breath holds (starting with the exhale and finishing at the end of the breath hold) are designated with the shaded regions. The SpO\textsubscript{2} recorded from the right index finger is denoted by the solid blue line, and the SpO\textsubscript{2} recorded from the right ear canal by the solid red line. The three significant drops in SpO\textsubscript{2} levels correspond to the three breath holds. The breath holds also vary slightly in length as they adhere to when the subject holds down a button rather than the set time of the instructional display, thus mirroring the true breath hold duration more accurately.}
\label{example_recording}
\end{figure*}

The participants in the recordings were 14 healthy subjects (7 males, 7 females) aged 19 - 38 years. Two PPG sensors were used per subject, one secured within the right ear canal and the other to the right index finger; both sensors recorded simultaneously. The subjects were seated in front of a monitor during the recording where a video guided them on when to breathe normally, as well as when to exhale and hold their breath. The position of the sensors, monitor and recording device relative to the subject is shown in Fig.~\ref{recording_diagram}. This video included a built in count down, so that at every stage of the protocol, depicted in Fig.~\ref{flow_digram_experiment}, the subject knew how long they had left. The experiment lasted for 435 seconds, consisting of 120 seconds of normal breathing, 3 repeats of 5 seconds of exhaling and 20 seconds of breath holding with 60 seconds of normal breathing between, and a final 120 seconds of normal breathing, as depicted in Fig.~\ref{flow_digram_experiment}. The exhale before the breath hold was included because if the oxygen from the lungs is expelled before holding one's breath, there is a sharper desaturation in blood oxygen than if a subject did not exhale first, thus allowing for clearer comparisons of sensitivity between the different measurement sites. Furthermore, in a real world scenario, such as obstructive sleep apnea, the lungs would not be filled with oxygen before the obstruction to inspiration occurred, making the conventional breath hold, where a subject inhales before holding their breath, less realistic. Subjects were also told to hold down a button, connected to the circuit board, from the moment they started exhaling until they stopped the breath hold. This was necessary, in order to know precisely when the subject was holding their breath, as reaction times would add uncertainty if the instructional display was used as the ground truth. The recordings were performed under the 881/MODREC/18 ethics approval, and all subjects gave full informed consent.

\subsection{Extraction of the SpO\textsubscript{2} Signal}

\begin{figure}[h]
\centerline{\includegraphics[width=0.5\textwidth]{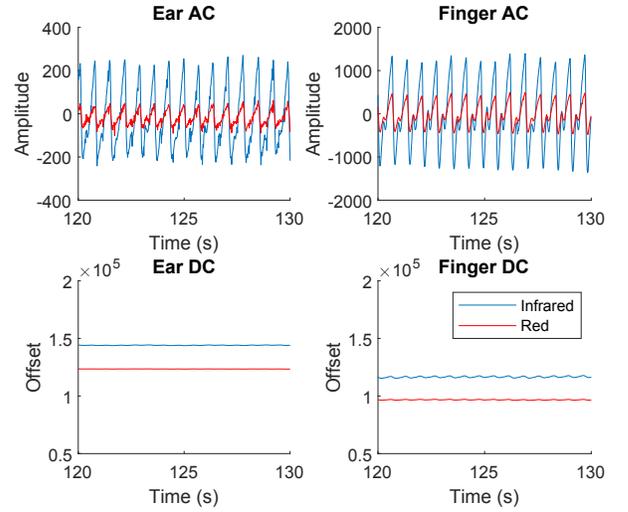}}
\caption{Examples of the alternating current (AC) and direct current (DC) waveforms within a PPG measurement, from both the right ear canal and the right index finger. The amplitude is lower in the right ear canal for both the red and infrared signals, but the signal is clear and usable. The maximum to minimum amplitude was used as the AC signal and is divided by the DC signal in the ratio of ratios metric to calculate SpO\textsubscript{2}.}
\label{ppg_example_waveform}
\end{figure}

The ratio of absorbance of infra-red to red light within the PPG sensor changes depending on the proportion of haemoglobin that is oxygenated in the blood. This change can be quantified through the so called ratio of ratios metric \cite{Rusch1996}, given by

\begin{equation}
    R = \frac{\frac{AC_{red}}{DC_{red}}}{\frac{AC_{infrared}}{DC_{infrared}}}
\end{equation}

An empirically derived linear approximation can then be used to calculate an SpO\textsubscript{2} value as a proxy to oxygen saturation. Using the manufacturers suggested calibration \cite{Maximintegrated2018}, the SpO\textsubscript{2} value was calculated as

\begin{equation}
    SpO_{2} = 104 - 17R
\end{equation}

To obtain the alternating current (AC) components within the PPG measurements, the raw signals were firstly band-pass filtered between 1Hz and 30Hz. Peak detection was then performed on the infrared and red AC filtered signals to find their peaks and troughs. Next, the peak values and trough values were separated and interpolated, before their absolute values were added together to give a constant estimate of the AC amplitude. The direct current (DC) components were obtained by low-pass filtering the raw signals at 0.01Hz. These waveforms are shown in Fig.~\ref{ppg_example_waveform}. 

\begin{figure*}[t]
\centerline{\includegraphics[width=\textwidth]{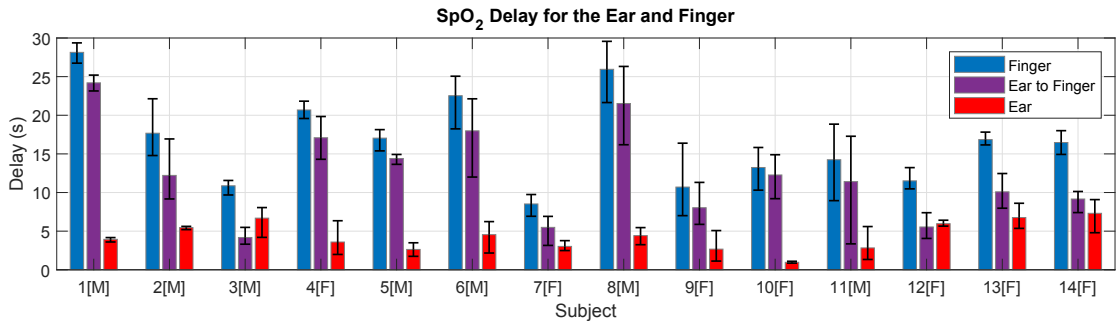}}
\caption{The absolute blood oxygen delay from the button press to the finger (blue), to the ear (red) and the relative delay between the ear and the finger (purple), shown for every subject. Male subjects are designated with [M] and female subjects with [F]. The delay is calculated from the time that the breath hold stops (true minimum blood oxygen) to the time at which the measuring site detects a minimum in SpO\textsubscript{2}. Error bars are included to show the range of the three delay measurements for individual subjects.}
\label{full_delay_subjects}
\end{figure*}

\subsection{Data Analysis}

\subsubsection{Resting SpO\textsubscript{2} comparison}
The resting SpO\textsubscript{2} was calculated over the 60 second section of normal breathing before the first exhale began. This time was ascertained from the timing of the first button press, explained in the Experimental Protocol subsection. The SpO\textsubscript{2} signal from the finger and ear were averaged individually over this 60 second window to give a resting blood oxygen comparison for each of the measuring sites for every subject. The root mean square difference in average resting SpO\textsubscript{2} was then calculated across all subjects, as well as the mean difference across all subjects.
\subsubsection{SpO\textsubscript{2} delay}
The blood oxygen estimation delay was calculated by using the button release point as the marker for minimal blood oxygen, corresponding to the point at which the breath hold ends. The time between this point of minimal blood oxygen and the first trough of the SpO\textsubscript{2} waveform for the ear and the finger was then used to calculate the SpO\textsubscript{2} delay for the ear, the finger and then the relative delay between the ear and the finger. Three measurements for delay were taken for each measuring site, one for every consecutive breath hold as shown in Fig.~\ref{example_recording}. The mean of these three recorded delays was taken to give an average delay for each measurement site for each subject. The range, on a per subject basis, was taken as a measure of variability within the recordings for each subject. The distribution of mean delays was analysed against age and sex, whereby a paired sample \textit{t}-test was employed to compare male and female delays while Pearson's correlation coefficient was utilised to determine if subject age was correlated to delay.

\section{Results}

For rigour, and to demonstrate non-inferiority of in-ear pulse oximetry to finger pulse oximetry, the experimental results span three major aspects related to the feasibility of in-ear PPG; these are the comparison of resting oxygen levels against standard finger PPG, the respective delays in detection of hypoxic events and the corresponding PPG signal amplitudes.


\subsection{Resting Oxygen Comparison}

\begin{figure}[h]
\centerline{\includegraphics[width=0.5\textwidth]{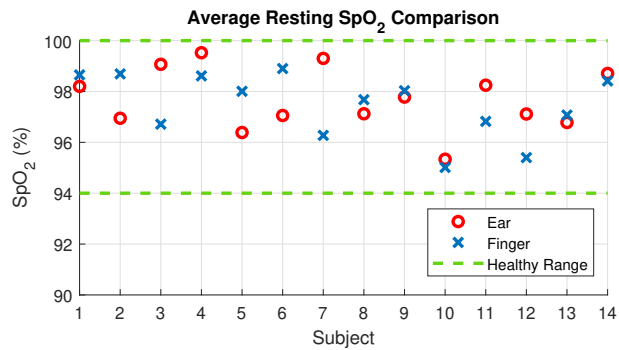}}
\caption{Mean resting SpO\textsubscript{2} levels taken for each individual subject for the right ear canal (red circles) and the right index finger (blue crosses). The mean was taken across the 60 seconds of SpO\textsubscript{2} data before the first breath hold sequence. The area within the two green dotted lines represents the healthy blood oxygen range of 94-100\%.}
\label{rest_o2}
\end{figure}

The difference in resting SpO\textsubscript{2} between the right ear canal and the right index finger across all subjects had a root mean square value of 1.47\% with a mean difference of 0.23\% higher saturation in the ear. The distribution of resting values across subjects is provided in Fig.~\ref{rest_o2}, and shows that all resting SpO\textsubscript{2} values were within the physiologically healthy range of 94-100\%, and therefore there was a complete agreement between the ear canal and finger as measuring sites on whether or not a subject had 'healthy' blood oxygen levels.

\subsection{Blood Oxygen Delay}

\begin{figure}[h]
\centerline{\includegraphics[width=0.42\textwidth]{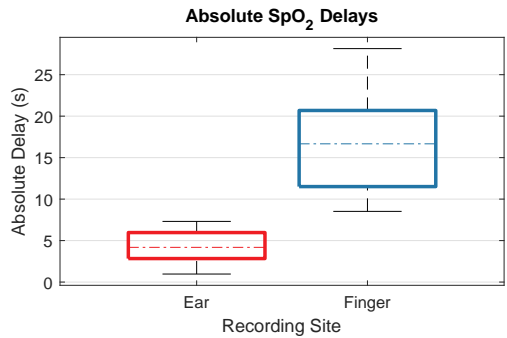}}
\caption{Boxplots of the mean absolute delays, for all subjects, between the right ear canal and the right index finger. The top and bottom of each coloured box represent respectively the upper and lower quartiles, the dotted lines represent the median, and the whisker lines extending out of the box represent the range. In our data, there is no overlap between the range of absolute delay from the ear canal and the absolute delay from the right index finger.}
\label{absolute_boxplot}
\end{figure}

The mean relative time delay per subject between the ear and finger pulse oximetry, in other words the time it took from detecting minimal blood oxygen in the ear to detecting minimal blood oxygen in the finger, ranged from 4.18 seconds to 24.2 seconds. The mean relative delay across all 14 subjects was 12.4 seconds, with a mean relative delay of 9.67 seconds for the female subjects and a mean relative delay of 15.13 seconds for the male subjects. A paired sample \textit{t}-test showed that the male and female means were significantly different (\textit{P}=.045). The SpO\textsubscript{2} delay values for all subjects are shown in Fig.~\ref{full_delay_subjects}, highlighting the large inter-subject variability in SpO\textsubscript{2} delay and moreover the large inter-trial variability in the ear canal vs finger delay for many individual subjects.

Across our 14 subjects there was no overlap between the absolute SpO\textsubscript{2} delay from the right ear canal and the right index finger, as shown in Figure~\ref{absolute_boxplot}. The absolute delay from the right ear canal had a mean of 4.35 seconds, with a range of 0.97-7.31 seconds. The absolute delay from the right index finger had a mean of 16.75 seconds, with a range of 8.52-28.14 seconds. When separated by sex, the mean absolute delay from the finger was 14.0 seconds for the female subjects and 19.5 seconds for the male subjects. The mean absolute delay from the ear canal was 4.3 seconds for females and 4.4 seconds for males, as summarised in Table~\ref{mean_delay_table}. Furthermore, there was no correlation found between age and oxygen delay across the participants.

For rigour, the SpO\textsubscript{2} delay between the left and right ear canal was also examined for a single subject, giving a mean delay of 0.46 seconds from the left ear canal to the right ear canal across the three breath holds, suggesting that measuring SpO\textsubscript{2} from the left ear would yield a similar response time to the right ear, if not slightly faster.

\begin{table}[ht] 
\caption{Summary of mean SpO\textsubscript{2} delay values (seconds)} 
\centering 
\begin{tabular}{c| ccc} 
\hline\hline 
  &Relative & Finger & Ear Canal \\ [0.25ex] 
\hline   
Female & 9.70 & 14.00 & 4.33  \\  
Male & 15.13 & 19.49 & 4.36  \\ 
 Total & 12.40 & 16.75 & 4.35  \\  
\hline\hline  
\end{tabular} 
\label{mean_delay_table} 
\end{table}

\subsection{Photoplethysmogram Amplitude}

\begin{figure}[h]
\centerline{\includegraphics[width=0.5\textwidth]{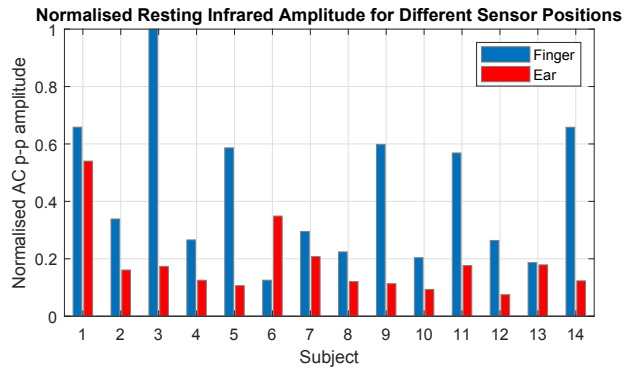}}
\caption{The mean resting maximum to minimum AC amplitude of the photoplethysmography signal for all subjects. These values were measured from the 60 seconds of filtered PPG data before the first breath hold and were normalised to the maximum mean AC amplitude across all subjects.}
\label{ppg_amplitude}
\end{figure}

It was observed that 13 out of 14 subjects had a higher AC PPG amplitude from the right index finger than the right ear canal with the mean finger PPG amplitude 2.35 times higher than for the ear canal. Fig.~\ref{ppg_amplitude} presents the individual normalised resting infrared PPG amplitude for the finger and ear canal results for all 14 subjects. Importantly, even in the lowest amplitude cases, the in-ear PPG signals were effective for peak detection and thus usable for the calculation of SpO\textsubscript{2}.

\section{Discussion}

\subsection{Resting SpO\textsubscript{2}}
Blood oxygen saturation measurements from the right ear canal and the right index finger were comparable at rest, with a minimal systematic offset between the two, indicated by the mean difference of 0.23\% higher resting SpO\textsubscript{2} in the right ear canal, as shown in Fig.~\ref{rest_o2}. The variability between the two measurement sites, indicated by the root mean square difference of 1.47\%, is to be expected, given that mean SpO\textsubscript{2} values were shown to vary as much as 0.9\% even across fingers on the same hand \cite{Basaranoglu2015}. The blood oxygen saturation delay between the ear and the finger may also be a source of error when using the same absolute 60 second time window for comparison, although given that the window was long in comparison with the SpO\textsubscript{2} delay and the fact that resting values stay fairly constant, this source of error should be minimal. Both sensor positions were in 100\% agreement concerning the patients having healthy SpO\textsubscript{2} levels of 94\% to 100\%. The established similarity in the resting oxygen saturation values with the commonly used right index finger location are a first step towards validating the ear canal as an alternative measurement site for pulse oximetry.

\subsection{SpO\textsubscript{2} Delay}
Our results indicate a mean oxygen saturation delay between the right ear canal and the right index finger of 12.4 seconds. In other words, a steep drop in SpO\textsubscript{2} is detected on average 12.4 seconds faster from the right ear canal than it is from the right index finger, with two subjects exhibiting over 20 seconds faster detection from the right ear canal. The faster response from the ear-canal strongly recommends the incorporation of in-ear SpO\textsubscript{2} measurements into so-called Hearables, as there are many situations where fast SpO\textsubscript{2} response time is critical, such as improving outcomes during neonatal resuscitation \cite{Baquero2011} and closed loop automatic oxygen therapy \cite{Morozoff2009}. Moreover, this opens the avenue to aid comparison of sleep stage scoring and sleep apnea events by aligning changes in SpO\textsubscript{2} to the correct 30 second sleep epoch \cite{Berry2017}.

We found a large inter-subject variability in both the absolute oxygen delay to the right ear canal and the absolute oxygen delay to the right index finger. On average, the female subjects had lower relative oxygen delay between the ear and the finger, caused by the much smaller absolute delay to the right index finger in combination with an almost identical absolute delay to the right ear canal. We hypothesise that this could be related to males being taller on average and therefore having longer arms, which would give oxygenated blood further to travel from the heart to the fingers. However, there may be other factors contributing to this large delay, as pulse transit times (the time for blood to travel from the heart) to the fingers are on the order of a few 100 milliseconds \cite{Fung2004} and not multiple seconds. Importantly, age was not found to be a correlating factor to oxygen delay in our subjects. In future, it would be important to investigate the SpO\textsubscript{2} delay in older cohorts that may have underlying cardiovascular conditions, and not just young healthy subjects. The natural question arises, could SpO\textsubscript{2} delay be an indicator of the health of a patients circulatory system?

\subsection{Photoplethysmogram Amplitude}
A significant drawback of in-ear PPG in this study was the far smaller amplitudes from the ear canal when compared with the finger. The two reasons for a lower PPG amplitude are: i) less vascular density of the tissue and ii) poorer sensor contact with the skin. Considering that the ear and fingers are both common PPG placement sites, given the high density of arterioles and capillaries \cite{Mannheimer2007}, the most likely explanation for a lower PPG amplitude was therefore sensor placement issues within the ear canal. The large variability in ear canal sizes was somewhat mitigated by different available earbud sizes, but even with the smallest earbud some subjects found it difficult to insert the sensor fully into the ear canal. Some subjects with wider ear canals also noted that the sensor became looser during the trial. These issues do not affect the conclusive nature of this feasibility study and will be addressed in future with an improved sensor design, such as by employing ear-hooks, commonly used with sports headphones, to stabilise the position of sensor within the ear. Importantly, despite a lower signal amplitude, all data recorded from the ear canal was functional for peak detection of the AC PPG signal and therefore the SpO\textsubscript{2} calculation.

\section{Conclusion}

\begin{table}[ht] 
\caption{Summary of different SpO\textsubscript{2} sensor positions} 
\centering 
\begin{tabular}{c| ccc} 
\hline\hline 
Placement &Fast response& Robust to temperature\\ [0.25ex] 
\hline   
Finger & No & No\\  
Earlobe & Yes \cite{Hamber1999} & No\\ 
 \textbf{Ear canal} & \textbf{Yes} & \textbf{Yes} \cite{Budidha2015}\\  
\hline\hline  
\end{tabular} 
\label{summarytable} 
\end{table} 

Improved non-intrusive wearable SpO\textsubscript{2} monitoring is desirable particularly in COVID-19 outpatients with a threat of respiratory deterioration. To this end, we have investigated the possibility of estimation of blood oxygen saturation from the ear canal and have comprehensively demonstrated its non-inferiority compared with conventionally used finger pulse oximetry, with a significantly faster response time averaging 12.4 seconds. The results have indicated that the favourable speed of in-ear pulse oximetry, in conjunction with the previously documented advantage of resilience to changes in circulation associated with environmental temperature changes, may offer significant clinical advantages. Given its privileged position on the human body and a fixed distance to vital signs during most daily activities and while sleeping, the ear canal may even be a superior site for measurement of SpO\textsubscript{2} in the scenario of 24/7 monitoring. In an era of urgency for enhanced tracking of COVID-19 symptoms and increasing use of smart devices for health monitoring, this study provides a compelling argument for the integration of ear canal pulse oximetry into current state-of-the-art Hearables. 

\section*{Acknowledgment}
This work was supported by the Racing Foundation grant 285/2018, MURI/EPSRC grant EP/P008461, the Dementia Research Institute at Imperial College London, the BHF Centre of Research Excellence, the Imperial Centre for Cardiac Engineering, the Imperial Health Charity and the NIHR Biomedical Research Centre.

\FloatBarrier
\bibliographystyle{IEEEtran}
\bibliography{IEEEabrv,bloodo2}

\begin{thebibliography}{10}
\providecommand{\url}[1]{#1}
\csname url@samestyle\endcsname
\providecommand{\newblock}{\relax}
\providecommand{\bibinfo}[2]{#2}
\providecommand{\BIBentrySTDinterwordspacing}{\spaceskip=0pt\relax}
\providecommand{\BIBentryALTinterwordstretchfactor}{4}
\providecommand{\BIBentryALTinterwordspacing}{\spaceskip=\fontdimen2\font plus
\BIBentryALTinterwordstretchfactor\fontdimen3\font minus
  \fontdimen4\font\relax}
\providecommand{\BIBforeignlanguage}[2]{{%
\expandafter\ifx\csname l@#1\endcsname\relax
\typeout{** WARNING: IEEEtran.bst: No hyphenation pattern has been}%
\typeout{** loaded for the language `#1'. Using the pattern for}%
\typeout{** the default language instead.}%
\else
\language=\csname l@#1\endcsname
\fi
#2}}
\providecommand{\BIBdecl}{\relax}
\BIBdecl

\bibitem{Jubran1999}
A.~Jubran, ``{Pulse oximetry},'' p. R11, May 1999.

\bibitem{Smith2012}
G.~B. Smith, D.~R. Prytherch, D.~Watson, V.~Forde, A.~Windsor, P.~E. Schmidt,
  P.~I. Featherstone, B.~Higgins, and P.~Meredith, ``{SpO2 values in acute
  medical admissions breathing air-Implications for the British Thoracic
  Society guideline for emergency oxygen use in adult patients?}''
  \emph{Resuscitation}, vol.~83, no.~10, pp. 1201--1205, Oct 2012.

\bibitem{WorldHealthOrganization2011}
{World Health Organization}, \emph{{Pulse Oximetry Training Manual}}, 2011.

\bibitem{Gattinoni2020a}
L.~Gattinoni, D.~Chiumello, P.~Caironi, M.~Busana, F.~Romitti, L.~Brazzi, and
  L.~Camporota, ``{COVID-19 pneumonia: different respiratory treatments for
  different phenotypes?}'' \emph{Intensive Care Med}, Apr 2020.

\bibitem{Cascella2020}
M.~Cascella, M.~Rajnik, A.~Cuomo, S.~C. Dulebohn, and R.~{Di Napoli},
  \emph{{Features, Evaluation and Treatment Coronavirus (COVID-19)}}.\hskip 1em
  plus 0.5em minus 0.4em\relax StatPearls Publishing, Apr 2020.

\bibitem{Pan2020}
F.~Pan, T.~Ye, P.~Sun, S.~Gui, B.~Liang, L.~Li, D.~Zheng, J.~Wang, R.~L.
  Hesketh, L.~Yang, and C.~Zheng, ``{Time Course of Lung Changes On Chest CT
  During Recovery From 2019 Novel Coronavirus (COVID-19) Pneumonia},''
  \emph{Radiology}, p. 200370, Feb 2020.

\bibitem{Alhazzani2020}
W.~Alhazzani, M.~H. M{\o}ller, Y.~M. Arabi, M.~Loeb, M.~N. Gong \emph{et~al.},
  ``{Surviving Sepsis Campaign: guidelines on the management of critically ill
  adults with Coronavirus Disease 2019 (COVID-19)},'' 2020.

\bibitem{Khemani2012}
R.~G. Khemani, N.~J. Thomas, V.~Venkatachalam, J.~P. Scimeme, T.~Berutti, J.~B.
  Schneider, P.~A. Ross, D.~F. Willson, M.~W. Hall, and C.~J.~L. Newth,
  ``{Comparison of SpO2 to PaO2 based markers of lung disease severity for
  children with acute lung injury},'' \emph{Critical Care Medicine}, vol.~40,
  no.~4, pp. 1309--1316, Apr 2012.

\bibitem{Oksenberg2010}
A.~Oksenberg, E.~Arons, K.~Nasser, T.~Vander, and H.~Radwan, ``{REM-related
  Obstructive Sleep Apnea: The Effect of Body Position},'' Tech. Rep.~4, 2010.

\bibitem{Nitzan2014}
M.~Nitzan, A.~Romem, and R.~Koppel, ``{Pulse oximetry: Fundamentals and
  technology update},'' pp. 231--239, Jul 2014.

\bibitem{Nilsson2007}
L.~Nilsson, T.~Goscinski, S.~Kalman, L.-G. Lindberg, and A.~Johansson,
  ``{Combined photoplethysmographic monitoring of respiration rate and pulse: a
  comparison between different measurement sites in spontaneously breathing
  subjects},'' \emph{Acta Anaesthesiologica Scandinavica}, vol.~51, no.~9, pp.
  1250--1257, Aug 2007.

\bibitem{Goverdovsky2017}
V.~Goverdovsky, W.~{Von Rosenberg}, T.~Nakamura, D.~Looney, D.~J. Sharp,
  C.~Papavassiliou, M.~J. Morrell, and D.~P. Mandic, ``{Hearables: Multimodal
  physiological in-ear sensing},'' \emph{Scientific Reports}, vol.~7, no.~1,
  pp. 1--10, Dec 2017.

\bibitem{Looney2012}
D.~Looney, P.~Kidmose, C.~Park, M.~Ungstrup, M.~Rank, K.~Rosenkranz, and
  D.~Mandic, ``{The in-the-ear recording concept: User-centered and wearable
  brain monitoring},'' \emph{IEEE Pulse}, vol.~3, no.~6, pp. 32--42, 2012.

\bibitem{Nakamura2020}
T.~Nakamura, Y.~D. Alqurashi, M.~J. Morrell, and D.~P. Mandic, ``{Hearables:
  Automatic Overnight Sleep Monitoring with Standardized In-Ear EEG Sensor},''
  \emph{IEEE Transactions on Biomedical Engineering}, vol.~67, no.~1, pp.
  203--212, Jan 2020.

\bibitem{Budidha2015}
K.~Budidha and P.~A. Kyriacou, ``{Investigation of photoplethysmography and
  arterial blood oxygen saturation from the ear-canal and the finger under
  conditions of artificially induced hypothermia},'' in \emph{Proceedings of
  the IEEE Annual International Conference of the Engineering in Medicine and
  Biology Society, EMBS}, Nov 2015, pp. 7954--7957.

\bibitem{Budidha2018}
------, ``{In vivo investigation of ear canal pulse oximetry during
  hypothermia},'' \emph{Journal of Clinical Monitoring and Computing}, vol.~32,
  no.~1, pp. 97--107, Feb 2018.

\bibitem{Hamber1999}
E.~A. Hamber, P.~L. Bailey, S.~W. James, D.~T. Wells, J.~K. Lu, and N.~L. Pace,
  ``{Delays in the detection of hypoxemia due to site of pulse oximetry probe
  placement},'' \emph{Journal of Clinical Anesthesia}, vol.~11, no.~2, pp.
  113--118, Mar 1999.

\bibitem{Goverdovsky2016}
V.~Goverdovsky, D.~Looney, P.~Kidmose, and D.~P. Mandic, ``{In-Ear EEG from
  viscoelastic generic earpieces: robust and unobtrusive 24/7 monitoring},''
  \emph{IEEE Sensors Journal}, vol.~16, no.~1, pp. 271--277, Jan 2016.

\bibitem{Rusch1996}
T.~L. Rusch, R.~Sankar, and J.~E. Scharf, ``{Signal processing methods for
  pulse oximetry},'' \emph{Computers in Biology and Medicine}, vol.~26, no.~2,
  pp. 143--159, 1996.

\bibitem{Maximintegrated2018}
MaximIntegrated, ``{Recommended Configurations and Operating Profiles for
  MAX30101/MAX30102 EV Kits},'' Tech. Rep., 2018.

\bibitem{Basaranoglu2015}
G.~Basaranoglu, M.~Bakan, T.~Umutoglu, S.~U. Zengin, K.~Idin, and Z.~Salihoglu,
  ``{Comparison of SpO2 values from different fingers of the hands},''
  \emph{SpringerPlus}, vol.~4, no.~1, Dec 2015.

\bibitem{Baquero2011}
H.~Baquero, R.~Alviz, A.~Castillo, F.~Neira, and A.~Sola, ``{Avoiding
  hyperoxemia during neonatal resuscitation: time to response of different SpO2
  monitors},'' \emph{Acta Paediatrica}, vol. 100, no.~4, pp. 515--518, Apr
  2011.

\bibitem{Morozoff2009}
E.~P. Morozoff and J.~A. Smyth, ``{Evaluation of three automatic oxygen therapy
  control algorithms on ventilated low birth weight neonates},'' in
  \emph{Proceedings of the 31st IEEE Annual International Conference of the
  Engineering in Medicine and Biology Society}, 2009, pp. 3079--3082.

\bibitem{Berry2017}
R.~B. Berry, R.~Brooks, C.~E. Gamaldo, S.~M. Harding, R.~M. Lloyd, S.~F. Quan,
  M.~M. Troester, and B.~V. Vaughn, ``{AASM | Scoring Manual Version 2.4 The
  AASM Manual for the Scoring of Sleep and Associated Events RULES, TERMINOLOGY
  AND TECHNICAL SPECIFICATIONS VERSION 2.4},'' Tech. Rep., 2017.

\bibitem{Fung2004}
P.~Fung, G.~Dumont, C.~Ries, C.~Mott, and M.~Ansermino, ``{Continuous
  noninvasive blood pressure measurement by pulse transit time},'' in
  \emph{Proceedings of the 26th IEEE Annual International Conference of the
  Engineering in Medicine and Biology Society}, vol. 26 I, 2004, pp. 738--741.

\bibitem{Mannheimer2007}
P.~D. Mannheimer, ``{The Light–Tissue Interaction of Pulse Oximetry},''
  \emph{Anesthesia {\&} Analgesia}, vol. 105, no. On Line Suppl., pp. S10--S17,
  Dec 2007.

\end{thebibliography}

\end{document}